\documentclass[aps,prd,twocolumn,nofootinbib]{revtex4}

\usepackage{graphicx}
\usepackage{amssymb}
\usepackage{color}
\usepackage[colorlinks=true,linkcolor=blue,citecolor=blue, urlcolor=blue]{hyperref}

\begin{document}

\title{$Z$-boson hadronic decay width up to ${\cal O}(\alpha_s^4)$-order QCD corrections using the single-scale approach of the principle of maximum conformality}

\author{Xu-Dong Huang$^1$}
\email{hxud@cqu.edu.cn}

\author{Xing-Gang Wu$^1$}
\email{wuxg@cqu.edu.cn (Corresponding author)}

\author{Xu-Chang Zheng$^1$}
\email{zhengxc@cqu.edu.cn}

\author{Qing Yu$^1$}
\email{yuq@cqu.edu.cn}

\author{Sheng-Quan Wang$^2$}
\email{sqwang@cqu.edu.cn}

\author{Jian-Ming Shen$^3$}
\email{shenjm@hnu.edu.cn}

\affiliation{$^1$ Department of Physics, Chongqing University, Chongqing 401331, People's Republic of China}
\affiliation{$^2$ Department of Physics, Guizhou Minzu University, Guiyang 550025, People’s Republic of China}
\affiliation{$^3$ School of Physics and Electronics, Hunan University, Changsha 410082, People's Republic of China}

\date{\today}

\begin{abstract}

In the paper, we study the properties of the $Z$-boson hadronic decay width by using the $\mathcal{O}(\alpha_s^4)$-order quantum chromodynamics (QCD) corrections with the help of the principle of maximum conformality (PMC). By using the PMC single-scale approach, we obtain an accurate renormalization scale-and-scheme independent perturbative QCD (pQCD) correction for the $Z$-boson hadronic decay width, which is independent to any choice of renormalization scale. After applying the PMC, a more convergent pQCD series has been obtained; and the contributions from the unknown $\mathcal{O}(\alpha_s^5)$-order terms are highly suppressed, e.g. conservatively, we have $\Delta \Gamma_{\rm Z}^{\rm had}|^{{\cal O}(\alpha_s^5)}_{\rm PMC}\simeq \pm 0.004$ MeV. In combination with the known electro-weak (EW) corrections, QED corrections, EW-QCD mixed corrections, and QED-QCD mixed corrections, our final prediction of the hadronic $Z$ decay width is $\Gamma_{\rm Z}^{\rm had}=1744.439^{+1.390}_{-1.433}$ MeV, which agrees with the PDG global fit of experimental measurements, $1744.4\pm 2.0$ MeV.

\end{abstract}

%\pacs{14.70.Hp, 12.38.Bx, 11.15.Bt}

\maketitle

\section{Introduction}

In quantum chromodynamics (QCD), the $Z$-boson hadronic decay width plays an important role in determining the strong coupling constant $(\alpha_s)$. The $Z$-boson hadronic decay width has been measured by various collaborations at the electron-positron colliders such as LEP and SLC~\cite{ALEPH:2005ab, Alcaraz:2007ri}, which could also be precisely measured in future high luminosity colliders such as the super $Z$ factory~\cite{superZ} or CEPC~\cite{CEPCStudyGroup:2018ghi}. Theoretically, the one-loop electroweak (EW) and the mixed EW-QCD contributions to the $Z$-boson hadronic decay have been investigated in Refs.\cite{Akhundov:1985fc, Novikov:1992rj, Novikov:1993ir, Czarnecki:1996ei}, and the two-loop EW contribution has been given in Refs.\cite{Freitas:2014hra, Freitas:2013dpa, Dubovyk:2018rlg}. In large-$m_t$ limit, the higher-loop corrections have been calculated up to $\mathcal{O}(\alpha_t \alpha_s^2)$~\cite{Avdeev:1994db, Chetyrkin:1995ix}, $\mathcal{O}(\alpha_t^2 \alpha_s)$, $\mathcal{O}(\alpha_t^3)$~\cite{Faisst:2003px, vanderBij:2000cg}, and $\mathcal{O}(\alpha_t \alpha_s^3)$~\cite{Schroder:2005db, Chetyrkin:2006bj, Boughezal:2006xk}, respectively, where $\alpha_t\equiv y_t^2/(4\pi)$ with $y_t$ being the top-quark Yukawa coupling constant. The final-state QED radiations have been computed up to $\mathcal{O}(\alpha)$, $\mathcal{O}(\alpha \alpha_s)$, $\mathcal{O}(\alpha^2)$ in Ref.\cite{Kataev:1992dg}. The non-factorizable QCD corrections have been estimated in Refs.\cite{Czarnecki:1996ei, Harlander:1997zb}. The pure perturbative QCD (pQCD) corrections up to $\mathcal{O}(\alpha_s^2)$~\cite{Kniehl:1989bb, Kniehl:1989qu}, $\mathcal{O}(\alpha_s^3)$~\cite{Gorishnii:1990vf, Surguladze:1990tg, Chetyrkin:1993ug, Chetyrkin:1993jm, Larin:1993ju}, $\mathcal{O}(\alpha_s^4)$~\cite{Baikov:2008jh, Baikov:2012er, Baikov:2012xh} have also been performed in the literature. Moreover, the mass corrections to both the vector and axial vector correlators can be found in Refs.\cite{Chetyrkin:1994js, Chetyrkin:2000zk, Baikov:2004ku, Kniehl:1989qu, Kniehl:1989bb, Chetyrkin:1993tt, Larin:1994va}. Those achievements give us good opportunities for precise determining of $\alpha_s(M_Z)$, e.g. a recent determination of $\alpha_s$ has been given in Ref.\cite{DEnterria:2019lql}.

Following the renormalization group invariance, the physical observable should be independent to theoretical conventions, such as the choices of the renormalization scale and scheme, which is ensured by mutual cancellation of the scale and scheme dependence among different orders for an infinite-order pQCD prediction. However, for a fixed-order pQCD prediction, if the perturbative coefficients and the corresponding $\alpha_s$ do not match properly, the pQCD series may have large scale and scheme ambiguities~\cite{Brodsky:2012ms}. Conventionally, the renormalization scale is taken as the ``guessed" momentum flow of the process, as well as the one to eliminate the large logs or to minimize the contributions from high-order terms or to achieve the prediction in agreement with the experimental data. Those naive treatment directly causes the mismatching between the strong coupling constant and its coefficients and resulting in conventional renormalization scale and scheme ambiguities~\cite{Wu:2013ei, Wu:2014iba}. Such guessing treatment decreases the predictive power of pQCD. In fact, predictions based on conventional scale setting are even incorrect for Abelian theory -- Quantum Electrodynamics (QED); the renormalization scale of the QED coupling constant can be set unambiguously by using the Gell-Mann-Low method~\cite{GellMann:1954fq}.

A correct renormalization scale-setting approach is thus important for achieving an accurate fixed-order pQCD prediction. Many ways have been suggested in the literature, most of them such as the renormalization group improved effective coupling method (FAC)~\cite{Grunberg:1980ja, Grunberg:1982fw} and the principle of minimum sensitivity (PMS)~\cite{Stevenson:1980du, Stevenson:1981vj, Stevenson:1982wn} are designed to find an optimal renormalization scale of the process. On the contrary, the principle of maximum conformality (PMC)~\cite{Brodsky:2011ta, Brodsky:2011ig, Brodsky:2012rj, Mojaza:2012mf, Brodsky:2013vpa} provides a rigorous idea, whose purpose is not to find an optimal scale, but to determine the effective magnitude of $\alpha_s$ for a fixed-order pQCD series by using the renormalization group equation (RGE). The determined effective $\alpha_s$ is independent to any choice of renormalization scale, thus the conventional renormalization scale ambiguity is eliminated. Moreover, since all the scheme-dependent non-conformal $\{\beta_i\}$-terms have been eliminated, the resultant pQCD series becomes scheme independent conformal series; thus the conventional renormalization scheme ambiguity is simultaneously eliminated.

A recent demonstration of renormalization scale-and-scheme independence of PMC prediction has been given in Refs.\cite{Wu:2018cmb, Wu:2019mky}, where by using the $C$-scheme coupling~\cite{Boito:2016pwf}, it has been proven that the PMC prediction is independent of the choice of renormalization scale and scheme up to any fixed order. Generally, the convergence of the pQCD series can be improved due to the elimination of the divergent renormalon terms like $n!\beta_0^n\alpha_s^n$ or $n!\beta_0^n\alpha_s^{n+1}$~\cite{Beneke:1994qe, Neubert:1994vb, Beneke:1998ui}~\footnote{The $\beta_n(n\geq1)$-term could be estimated by using the approximation, $\beta_n\approx \beta_0^{n+1}$, which can be adopted for transforming the $\{\beta_i\}$-series at each order into $\beta_0$-power series.}.

The PMC accurate renormalization scale-and-scheme independent conformal series is helpful not only for achieving precise pQCD predictions but also for a reliable prediction on the contributions of unknown higher-orders; some applications can be found in Refs.\cite{Du:2018dma, Yu:2018hgw, Yu:2019mce, Huang:2020rtx, Yu:2020tri} which are estimated by using the $\rm Pad\acute{e}$ resummation approach~\cite{Basdevant:1972fe, Samuel:1992qg, Samuel:1995jc}. In the present paper, we shall adopt the PMC single-scale approach~\cite{Shen:2017pdu}~\footnote{In the original PMC multi-scale approach~\cite{Brodsky:2011ta, Mojaza:2012mf}, different types of $\{\beta_i\}$-terms are absorbed into $\alpha_s$ via an order-by-order manner, and distinct PMC scales are determined at each order. Such PMC multi-scale approach has thus two kinds of residual scale dependence due to the unknown perturbative terms, which has already been pointed out in year 2013~\cite{Zheng:2013uja}. It should be pointed out that the recently so-called ambiguities of PMC given in Ref.\cite{Chawdhry:2019uuv} are not the default of PMC, but the residual scale dependence due to unknown perturbative terms. Such residual scale dependence generally suffer from both the $\alpha_s$-power suppression and the exponential suppression, but could be large due to possibly poor pQCD convergence for the perturbative series of either the PMC scale or the pQCD approximant~\cite{Wu:2019mky}.} to analyze the $Z$-boson hadronic decay width. It is noted that the original PMC multi-scale approach~\cite{Brodsky:2011ta, Brodsky:2011ig, Brodsky:2012rj, Mojaza:2012mf, Brodsky:2013vpa} and single-scale approach are equivalent to each other in sense of perturbative theory~\cite{Shen:2017pdu}, but the residual scale dependence emerged in PMC multi-scale method can be greatly suppressed by applying the single-scale approach.

The remaining parts of the paper are organized as follows. In Sec.II, we will give the detailed PMC treatment for a precise determination of the $Z$-boson hadronic decay width. In Sec.III, we will give the numerical results. Sec.VI is reserved for a summary.

\section{The $Z$-boson hadronic decay width using the PMC}

The hadronic decay width of the $Z$-boson can be expressed as
\begin{eqnarray}
\Gamma_{\rm Z}^{\rm had}=\Gamma_0R^{\rm nc}+\Delta\Gamma_{\rm Z}^{\rm Extra}, \label{Gammazhad}
\end{eqnarray}
where the first term stands for the pure pQCD correction with the leading-order (LO) width $\Gamma_0=\frac{G_F M^3_Z}{24\pi\sqrt{2}}$, and the Fermi coupling constant $G_F=1.166\times 10^{-5} {\rm GeV}^{-2}$. The second term $\Delta\Gamma_{\rm Z}^{\rm Extra}$ contains four less important corrections, i.e.,
\begin{eqnarray}
\Delta\Gamma_{\rm Z}^{\rm Extra}&=&\Delta\Gamma_1+\Delta\Gamma_2+\Delta\Gamma_3+\Delta\Gamma_4 \nonumber\\
&=&-1.577^{+0.183}_{-0.237}+0.695^{+0.000}_{-0.001}+6.577^{+0.560}_{-0.560} \nonumber \\
&&+0.609^{+0.061}_{-0.049} ~({\rm MeV}) \nonumber \\
&=&6.304^{+0.804}_{-0.847} ~({\rm MeV}),   \label{extra}
\end{eqnarray}
where the central values are for $\mu_r=M_Z$, and the errors are for $\mu_r\in[M_Z/2, 2M_Z]$. Here $\Delta\Gamma_1$ is the $b$- and $t$- quark mass corrections to the vector and axial vector correlators~\cite{Chetyrkin:1994js, Chetyrkin:2000zk, Baikov:2004ku, Chetyrkin:1993tt, Larin:1994va}, $\Delta\Gamma_2$ is the quark final-state QED radiation and the mixed QED-QCD correction~\cite{Kataev:1992dg}, $\Delta\Gamma_3$ is the electro-weak two-loop corrections and the higher-loop corrections in the large-$m_t$ limit~\cite{Dubovyk:2018rlg}, $\Delta\Gamma_4$ is the mixed EW-QCD correction and nonfactorizable QCD correction~\cite{Novikov:1992rj, Novikov:1993ir, Czarnecki:1996ei, Harlander:1997zb}.

Our main concern is the perturbative QCD corrections to the dominant correlator of the neutral current, which can be divided as the following four parts:
\begin{eqnarray}
R^{\rm nc}=3[\sum_f v^2_f r^V_{\rm NS}+\big(\sum_f v_f\big)^2 r^V_ S+\sum_f a^2_f r^ A_{\rm NS}+r^A_S],
\end{eqnarray}
where $v_f\equiv 2I_f-4q_fs^2_W$, $a_f\equiv2I_f$, $q_f$ is the $f$-quark electric charge, $s_W$ is the effective weak mixing angle, and $I_f$ is the third component of weak isospin of the left-handed component of $f$. $r^V_{\rm NS}=r^A_{\rm NS}\equiv r_{\rm NS}$, $r^V_S$, and $r^A_S$ stand for the non-singlet, the vector-singlet, and the axial-singlet part, respectively. Those contributions can be further expressed as
\begin{displaymath}
r_{\rm NS} = 1+\sum^n_{i=1}C^{\rm NS}_ia^i_s, r^V_S =\sum^n_{i=3}C^{\rm VS}_{i}a^i_s, ~r^A_S=\sum^n_{i=2}C^{\rm AS}_{i}a^i_s ,
\end{displaymath}
where $a_s=\alpha_s/(4\pi)$, and the coefficients of $r_{\rm NS}$, $r^V_S$, and $r^A_S$ can be obtained from Refs.\cite{Baikov:2012xh, Baikov:2012er, Baikov:2008jh, Baikov:2012zn, Baikov:2010je}. As for $r^A_S$, we adopt conventional scale setting approach to perform our analysis~\footnote{From the known ${\cal O}(\alpha_s^4)$-order expressions, we cannot derive the exact RG-dependent $n_f$-series for $r^A_S$, which is however very important for using the PMC scale-setting; so we have to take this approximation.}, and numerically, we obtain $\Gamma_{\rm Z}^{\rm had}|^A_S=[-1.725,-1.685]$ MeV for $\mu_r \in [M_Z/2, 2M_Z]$ by using the formulas given by Ref.\cite{Baikov:2012xh}, whose magnitude is quite small in comparison to that of $r_{\rm NS}$, thus fortunately, this approximate treatment will not affect our final conclusions.

The R-ratio can be rewritten as the following perturbative form by using the degeneracy relations~\cite{Mojaza:2012mf, Brodsky:2013vpa, Bi:2015wea}, i.e.,
\begin{eqnarray}
R^{\rm nc}&=&r_0+r_{1,0}a_s(\mu_r) + (r_{2,0}+\beta_{0}r_{2,1})a_{s}^{2}(\mu_r)\nonumber\\
&&+(r_{3,0}+\beta_{1}r_{2,1}+ 2\beta_{0}r_{3,1}+ \beta_{0}^{2}r_{3,2})a_{s}^{3}(\mu_r)\nonumber\\
&& +(r_{4,0}+\beta_{2}r_{2,1}+ 2\beta_{1}r_{3,1} + \frac{5}{2}\beta_{1}\beta_{0}r_{3,2} \nonumber\\
&& +3\beta_{0}r_{4,1}+3\beta_{0}^{2}r_{4,2}+\beta_{0}^{3}r_{4,3}) a_{s}^{4}(\mu_r)+\mathcal{O}(a^5_s),~\label{rij}
\end{eqnarray}
where $r_0=3(\sum_f v^2_f+\sum_f a^2_f)$, and the coefficients $r_{i,j}$ can be obtained from the known coefficients $C^{\rm NS}$, $C^{\rm VS}$, and $C^{\rm AS}_{i}$ of $r_{\rm NS}$, $r^V_S$, and $r^A_S$. The coefficients $r_{i,0}$ are $\{\beta_i\}$-independent conformal coefficients, and the $\{\beta_i\}$-dependent non-conformal coefficients $r_{i,j}$ $(j\neq0)$ are generally functions of $\ln\mu^{2}_{r} /M_{Z}^{2}$, i.e.,
\begin{equation}
r_{i,j}=\sum^j_{k=0}C^k_j{\hat r}_{i-k,j-k}{\rm ln}^k(\mu_r^2/M_Z^2),    \label{rijrelation}
\end{equation}
where the reduced coefficients ${\hat r}_{i,j}=r_{i,j}|_{\mu_r=M_Z}$, the combination coefficients $C^k_j=j!/[k!(j-k)!]$. We put the known coefficients ${\hat r}_{i,j}$ up to ${\cal O}(\alpha_s^4)$-level in the Appendix.

Following the standard PMC single-scale procedures as described in detail in Ref.\cite{Shen:2017pdu}, with the help of RGE, one can determine an effective coupling $\alpha_s(Q_*)$ by absorbing all the non-conformal $\{\beta_i\}$-terms into the running coupling, and the resultant pQCD series becomes the following conformal series,
\begin{eqnarray}
R^{\rm nc}|_{\rm PMC}&=&r_0+r_{1,0}a_s(Q_*)+r_{2,0}a^2_s(Q_*)\nonumber \\
&&+r_{3,0}a^3_s(Q_*)+r_{4,0}a^4_s(Q_*)+\mathcal{O} (a^5_s),\label{conformal}
\end{eqnarray}
where $Q_*$ is the PMC scale, which corresponds to the overall effective momentum flow of the process and can be determined up to next-to-next-to-leading log (NNLL) accuracy by using the present known ${\cal O}(\alpha_s^4)$-order pQCD series; i.e., the $\ln{Q^2_*}/{M^2_Z}$ can be expanded as the following perturbative series,
\begin{eqnarray}
\ln\frac{Q^2_*}{M^2_Z}=T_0+T_1 a_s(M_Z)+T_2 a^2_s(M_Z)+ {\cal O}(a^3_s),
\label{qstar}
\end{eqnarray}
where
\begin{eqnarray}
T_0=&&-\frac{{\hat r}_{2,1}}{{\hat r}_{1,0}}, \label{Tij1} \\
T_1=&&\frac{ \beta _0 ({\hat r}_{2,1}^2-{\hat r}_{1,0} {\hat r}_{3,2})}{{\hat r}_{1,0}^2}+\frac{2 ({\hat r}_{2,0} {\hat r}_{2,1}-{\hat r}_{1,0} {\hat r}_{3,1})}{{\hat r}_{1,0}^2},\label{Tij2}
\end{eqnarray}
and
\begin{eqnarray}
T_2=&&\frac{3 \beta _1 ({\hat r}_{2,1}^2-{\hat r}_{1,0} {\hat r}_{3,2})}{2 {\hat r}_{1,0}^2}\nonumber\\
&&+\frac{4({\hat r}_{1,0} {\hat r}_{2,0} {\hat r}_{3,1}-{\hat r}_{2,0}^2 {\hat r}_{2,1})+3({\hat r}_{1,0} {\hat r}_{2,1} {\hat r}_{3,0}-{\hat r}_{1,0}^2 {\hat r}_{4,1})}{ {\hat r}_{1,0}^3} \nonumber \\
&&+\frac{ \beta _0  (4 {\hat r}_{2,1} {\hat r}_{3,1} {\hat r}_{1,0}-3 {\hat r}_{4,2} {\hat r}_{1,0}^2+2 {\hat r}_{2,0} {\hat r}_{3,2} {\hat r}_{1,0}-3 {\hat r}_{2,0} {\hat r}_{2,1}^2)}{ {\hat r}_{1,0}^3}\nonumber\\
&&+\frac{ \beta _0^2 (2 {\hat r}_{1,0} {\hat r}_{3,2} {\hat r}_{2,1}- {\hat r}_{2,1}^3- {\hat r}_{1,0}^2 {\hat r}_{4,3})}{ {\hat r}_{1,0}^3}. \label{Tij3}
\end{eqnarray}
It can be found that $Q_*$ is exactly free of $\mu_r$, and together with the $\mu_r$-independent conformal coefficients $r_{i,0}$, the conventional renormalization scale ambiguity is eliminated. Therefore, the precision of $R^{\rm nc}$ can be greatly improved by using the PMC. Moreover, the precision of the predictions depend on the perturbative nature of both the $R^{\rm nc}$ and the $\ln {Q^2_*} / {M^2_Z}$, which shall be numerically analyzed in the following paragraphs.

\section{Numerical results}

To do the numerical calculation, we adopt the $Z$-boson mass $M_{Z}=91.1876\pm0.0021$ GeV and top-quark pole mass $M_{t}=172.9$ GeV~\cite{Tanabashi:2018oca}. We use the four-loop $\alpha_s$-running behavior~\cite{Wu:2019mky} to analyse the ${\cal O}(\alpha_s^4)$-order QCD corrections. i.e.,
\begin{eqnarray}
\alpha_s(\mu_r)&\simeq&\frac{1}{\beta_0 t}-\frac{b_1\ln t}{(\beta_0 t)^2}+\frac{b_1^2(\ln^2 t-\ln t-1)+b_2}{(\beta_0 t)^3} \nonumber \\
&&+\frac{1}{(\beta_0 t)^4}\bigg[b_1^3\left(-\ln^3 t+\frac{5}{2}\ln^2 t+2\ln t-\frac{1}{2}\right) \nonumber \\
&&-3b_1 b_2\ln t+\frac{b_3}{2}\bigg]+\mathcal{O}\left(\frac{1}{(\beta_0 t)^5}\right), \nonumber
\end{eqnarray}
Where $t=\ln(\mu_r^2/\Lambda_{\rm QCD}^2)$, $b_i=\beta_i/\beta_0$, and the $\beta_i(i=0,1,2,3)$-functions have been calculated in Refs.~\cite{Gross:1973id, Politzer:1973fx, Caswell:1974gg, Tarasov:1980au, Larin:1993tp, vanRitbergen:1997va, Chetyrkin:2004mf, Czakon:2004bu, Baikov:2016tgj}. Taking $\alpha_s(M_{Z})=0.1181$~\cite{Tanabashi:2018oca}, we obtain $\Lambda^{(n_f=5)}_{\rm QCD}=209.5$ MeV.

\begin{figure}[htb]
\includegraphics[width=0.48\textwidth]{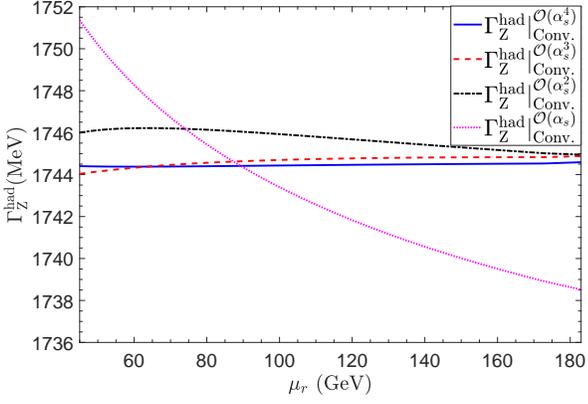}
\caption{The $Z$-boson hadronic decay width $\Gamma_{\rm Z}^{\rm had}$ under the conventional scale-setting approach versus the renormalization scale $\mu_r$. The dotted, the dashed-dot, the dashed and the solid lines are for the predictions up to ${\cal O}(\alpha_s)$, ${\cal O}(\alpha_s^2)$, ${\cal O}(\alpha_s^3)$, ${\cal O}(\alpha_s^4)$ order levels, respectively.}
\label{GammaConv}
\end{figure}

First, by setting all input parameters to be their central values, we present the $Z$-boson hadronic decay width $\Gamma_{\rm Z}^{\rm had}$ up to different known $\alpha_s$-orders under conventional scale-setting approach in Fig.~\ref{GammaConv}. It shows that in agreement of the conventional wisdom, the renormalization scale dependence becomes small when we have known more loop terms. For examples, we obtain $\Gamma_{\rm Z}^{\rm had}|_{\rm Conv.} =[1744.378,1744.587]$ MeV for $\mu_r \in [M_Z/2, 2M_Z]$, and $\Gamma_{\rm Z}^{\rm had}|_{\rm Conv.} =[1744.378,1745.008]$ MeV for $\mu_r \in [M_Z/3, 3M_Z]$; e.g., the net scale errors are only $\sim0.01\%$, and $\sim0.04\%$, respectively. We should point out that as has been mentioned in the Introduction, such small net scale dependence for the $\mathcal{O}(\alpha_s^4)$-order prediction is due to good convergence of the perturbative series, e.g., the relative magnitudes of the $\alpha_s$-terms: $\alpha_s^2$-terms: $\alpha_s^3$-terms: $\alpha_s^4$-terms=1: $2.9\%$: $-2.2\%$: $-0.4\%$ for the case of $\mu_r=M_Z$; and also due to the cancellation of the scale dependence among different orders~\footnote{In cases when each perturbative terms varies synchronously with the changes of $\mu_r$, there will have no cancellations of scale dependence among different orders and the pQCD approximant could be still large even for higher-orders. A recent example can be found in a two-loop QCD correction for $\gamma + \eta_c$ production in electron-positron collisions~\cite{Yu:2020tri}.}. The scale errors for each order term remain unchanged and large, e.g. the $\Gamma_{\rm Z}^{\rm had}$ has the following perturbative feature up to $\mathcal{O}(\alpha_s^4)$-order:
\begin{eqnarray}
\Gamma_{\rm Z}^{\rm had}|_{\rm Conv.}&=&1681.262+62.966^{-5.925}_{+4.268}+1.802^{+4.838}_{-4.078}\nonumber \\
&&-1.382^{+1.311}_{-0.505}-0.230^{-0.055}_{+0.275} \nonumber \\
&=&1744.418^{+0.169}_{-0.040}~({\rm MeV}),
\end{eqnarray}
where the central values are for $\mu_r=M_Z$, and the errors are obtained by varying $\mu_r \in [M_Z/2, 2M_Z]$. It shows that the absolute scale errors are $16\%$, $495\%$, $131\%$, and $143\%$ for the $\alpha_s$-terms, $\alpha_s^2$-terms, $\alpha_s^3$-terms, and $\alpha_s^4$-terms, respectively; and there do have large scale cancellations among different orders.

\begin{figure}[htb]
\includegraphics[width=0.48\textwidth]{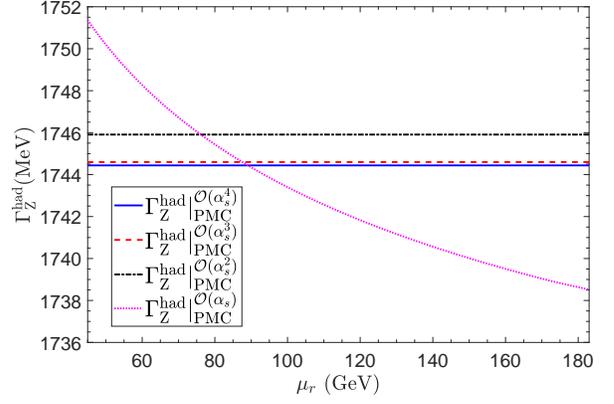}
\caption{The $Z$-boson hadronic decay width $\Gamma_{\rm Z}^{\rm had}$ under the PMC scale-setting approach versus the renormalization scale $\mu_r$. The dotted, the dashed-dot, the dashed and the solid lines are for the predictions up to ${\cal O}(\alpha_s)$, ${\cal O}(\alpha_s^2)$, ${\cal O}(\alpha_s^3)$, ${\cal O}(\alpha_s^4)$ order levels, respectively.}
\label{Gammapmc}
\end{figure}

Second, we present the $Z$-boson hadronic decay width $\Gamma_{\rm Z}^{\rm had}$ up to different known $\alpha_s$-orders under the PMC scale-setting approach in Fig.~\ref{Gammapmc}. At the $\mathcal{O}(\alpha_s)$-order level, the perturbative series of $\Gamma_{\rm Z}^{\rm had}$ does not have $\{\beta_i\}$-terms to fix the $\alpha_s$ value, thus the prediction of $\Gamma_{\rm Z}^{\rm had}|^{\mathcal{O}(\alpha_s)}_{\rm PMC}$ is the same as the conventional one. The PMC starts to work at the $\mathcal{O}(\alpha^2_s)$ and higher order levels. It shows that after applying the PMC, the pQCD convergence can be greatly improved, e.g., the relative magnitudes of the $\alpha_s$-terms: $\alpha_s^2$-terms: $\alpha_s^3$-terms of the pQCD series changes 1: $4.34\%$: $-0.49\%$ by applying the PMC scale-setting to the perturbative series up to $\mathcal{O}(\alpha_s^3)$, whose PMC scale $Q_*=113.0$ GeV is fixed up to the NLL accuracy. The relative magnitudes of the $\alpha_s$-terms: $\alpha_s^2$-terms: $\alpha_s^3$-terms: $\alpha_s^4$-terms of the pQCD series becomes to 1: $4.33\%$: $-0.49\%$: $0.01\%$ by applying the PMC up to $\mathcal{O}(\alpha_s^4)$, whose PMC scale $Q_*=114.9$ GeV is fixed up to NNLL accuracy. And there is no renormalization scale dependence for $\Gamma_{\rm Z}^{\rm had}$ at any fixed order, i.e.,
\begin{eqnarray}
\Gamma_{\rm Z}^{\rm had}|_{\rm PMC}&=&1681.262+60.838+2.634-0.299\nonumber \\
&&+0.004  \nonumber \\
&=&1744.439~({\rm MeV}),
\end{eqnarray}
where each perturbative terms and the net total decay width are unchanged for any choice of $\mu_r$. This behavior is consistent with that of the previous PMC multi-scale approach analysis on $R^{\rm nc}$~\cite{Wang:2014aqa}. The PMC single scale $Q_*$ is an effective scale which effectively replaces the individual PMC scales introduced in the PMC multi-scale approach in the sense of a mean value theorem, which can be regarded as the overall effective momentum flow of the process; it shows stability and convergence with increasing order in pQCD via the pQCD approximates. More explicitly, we obtain $Q_*=114.9$ GeV $\sim 1.3 M_Z$, which can be fixed up to NNLL accuracy by using the present known ${\cal O}(\alpha_s^4)$-order pQCD series, i.e.,
\begin{eqnarray}
\ln\frac{Q^2_*}{M^2_Z} &&=0.2249+21.7363a_s(M_Z)+376.287a^2_s(M_Z) \nonumber \\
&&=0.2249+0.2043+0.0332. \label{PMCsc}
\end{eqnarray}
One may observe that the relative magnitudes of each order terms in $Q_*$ perturbative series are $1: 91\%: 15\%$, which also shows a good convergence behavior.

Third, it is helpful to predict the magnitude of the ``unknown" higher-order pQCD corrections. The renormalization scale independent PMC series is helpful for such purpose. Because the PMC series has a good pertubative convergence, e.g., the magnitude of ${\cal O}(\alpha^4_s)$-order term is only $0.01\%$ of ${\cal O}(\alpha_s)$-order term, it is reasonable to take the magnitude of the last known term $\pm |r_{4,0} a^4_s(Q_*)|$ as a conservative prediction of the uncalculated higher-order terms~\cite{Wu:2014iba}. By further taking the variation of $\Delta Q_* \simeq \pm 1.9$ GeV, which is the difference between the NLL and NNLL PMC scales, as the magnitude of its unknown NNNLL term into consideration~\footnote{This is a conservative estimation of unknown contributions for the PMC scale $Q_*$, since as shown by Eq.(\ref{PMCsc}), $Q_*$ suffers from both the exponential suppression and $\alpha_s$ suppression, and it shows good convergence and is already at a high precision.}, we obtain
\begin{equation}
\Delta \Gamma_{\rm Z}^{\rm had}|^{\text{High order}}_{\rm PMC}\simeq \pm 0.004 ~({\rm MeV}).
\end{equation}

Finally, after eliminating the renormalization scale uncertainty by applying the PMC, we still have uncertainties from the $\alpha_s$ fixed-point error $\Delta\alpha_s(M_Z)$ and the $Z$-boson mass error $\Delta M_Z$. As for the $\alpha_s$ fixed-point error, by using $\Delta \alpha_s(M_Z) =0.0011$~\cite{Tanabashi:2018oca} together with the four-loop $\alpha_s$-running behavior, we obtain $\Lambda_{\rm QCD}^{n_f=5}=209.5^{+13.2}_{-12.6}$ MeV and
\begin{eqnarray}
\Delta\Gamma_{\rm Z}^{\rm had}|^{\Delta\alpha_s(M_Z)}_{\rm PMC}&&=\pm0.574 ~(\rm MeV).
\end{eqnarray}
And for the error of $Z$-boson mass $\Delta M_Z=\pm0.0021 \rm{GeV}$, we obtain
\begin{eqnarray}
\Delta\Gamma_{\rm Z}^{\rm had}|^{\Delta M_Z}_{\rm PMC}&&=\pm0.120 ~(\rm MeV).
\end{eqnarray}
Here, when discussing one uncertainty, the other input parameters shall be set as their central values.

As a whole, the squared average of the above mentioned three errors leads to a net error, $\pm 0.586$ MeV, to the PMC prediction of the total decay width $\Gamma_{\rm Z}^{\rm had}$, among which the magnitude of $\Delta\alpha_s(M_Z)$ dominates the error sources. Thus more precise measurements on the reference point $\alpha_s(M_Z)$ is important for a more precise pQCD prediction.

\section{Summary}

Under conventional scale-setting approach, the fixed-order scale-setting ambiguity could be softened by including enough higher-order loop terms due to large cancelation among different orders; for the present considered decay width up to ${\cal O}(\alpha_s^4)$-order, the net scale uncertainty is $\left(^{+0.169}_{-0.040}\right)$ MeV for $\mu_r \in [M_Z/2, 2M_Z]$; and by further including the mentioned other error sources, we have $\Gamma_{\rm Z}^{\rm had}|_{\rm Conv.}=1744.418^{+1.595}_{-1.621}$ (MeV).
\begin{figure}[htb]
\includegraphics[width=0.48\textwidth]{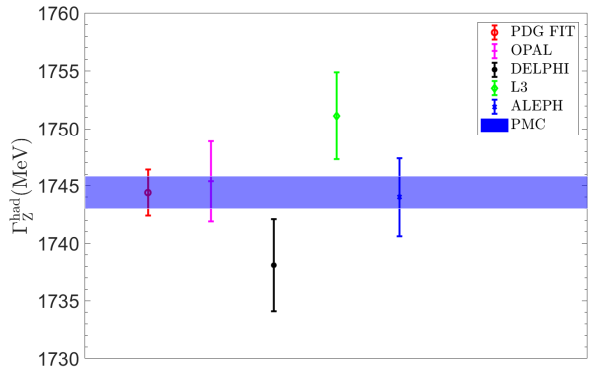}
\caption{The PMC prediction of the $Z$-boson hadronic decay in comparison with the experimental values given by the PDG global fit of experimental data~\cite{Tanabashi:2018oca}, and by the OPAL~\cite{Abbiendi:2000hu}, DELPHI~\cite{Abreu:2000mh}, L3~\cite{Acciarri:2000ai}, and ALEPH~\cite{Barate:1999ce} collaborations.}
\label{gammacon_pmc1}
\end{figure}

In the paper, we have presented an accurate prediction on the $Z$-boson hadronic decay width by applying the PMC single-scale approach to eliminate the conventional renormalization scale ambiguity. We obtain
\begin{eqnarray}
\Gamma_{\rm Z}^{\rm had}|_{\rm PMC} = 1744.439^{+1.390}_{-1.433} ~(\rm MeV).
\end{eqnarray}
where the errors are the sum of two parts, one is the squared average of those from $\Delta \alpha_s(M_Z)$, $\Delta M_Z$, and the uncalculated higher-order terms, another is the error from $\Delta\Gamma_{\rm Z}^{\rm Extra}$ as given in Eq.(\ref{extra}). After applying the PMC single-scale approach, the pQCD series becomes scale independent and more convergent, thus a reliable pQCD prediction can be achieved. Due to the perturbative terms have been known up to enough high-orders, the predictions under the PMC and conventional scale-setting approaches are consistent with each other. We present the PMC prediction of the $Z$-boson hadronic decay width in Fig.~\ref{gammacon_pmc1}, where the experimental data are presented as a comparison. The PMC prediction agrees with the PDG global fit of the experimental measurements. Thus, one obtains optimal fixed-order predictions for the $Z$-boson hadronic decay width by applying the PMC, enabling high precision test of the Standard Model.

\hspace{2cm}

{\bf Acknowledgements:} This work is partly supported by the Chongqing Graduate Research and Innovation Foundation under Grant No.CYB19065 and No.ydstd1912, the National Natural Science Foundation of China under Grant No.11625520, No.11905056, No.11975187, No.11947406, and the Fundamental Research Funds for the Central Universities under Grant No.2020CQJQY-Z003.

\appendix

\section*{Appendix: The PMC reduced perturbative coefficients ${\hat r}_{i,j}$}

In this appendix, we give the required PMC reduced coefficients ${\hat r}_{i,j}$ for the perturbative series of the $Z$-boson hadronic decay width, which can be obtained from Refs.\cite{Baikov:2012xh, Baikov:2012er, Baikov:2008jh, Baikov:2012zn, Baikov:2010je} with proper transformations, i.e.,
\begin{eqnarray}
{\hat r}_{1,0}&=& 9\gamma^{NS}_1(\sum_f a^2_f+\sum_f v^2_f), \\
{\hat r}_{2,0}&=& 4\left[9\gamma^{NS}_2(\sum_f a^2_f+\sum_f v^2_f)-37+12\ln{\frac{M_Z^2}{M_t^2}} \right], \\
{\hat r}_{2,1}&=& 9\Pi^{NS}_1(\sum_f a^2_f+\sum_f v^2_f), \\
{\hat r}_{3,0}&=& 144\left[\gamma^{NS}_3+\frac{\gamma^S_3\big(\sum_f q_f\big)^2}{\sum_f q^2_f}\right](\sum_f a^2_f+\sum_f v^2_f)\nonumber \\
&&+\big(\sum_f v_f\big)^2\left(\frac{440}{9}-\frac{320\zeta_3}{3}\right)+\frac{2144}{3}\ln{\frac{M_Z^2}{M_t^2}}\nonumber \\
&&+368\ln^2{\frac{M_Z^2}{M_t^2}}+192\zeta_3+\frac{368\pi^2}{3}-\frac{40600}{9}, \\
{\hat r}_{3,1}&=& 36\Pi^{NS}_2(\sum_f a^2_f+\sum_f v^2_f), \\
{\hat r}_{3,2}&=& -3\pi^2\gamma^{NS}_1(\sum_f a^2_f+\sum_f v^2_f),\\
{\hat r}_{4,0}&=& 576\left[\gamma^{NS}_4+\frac{\gamma^S_4\big(\sum_f q_f\big)^2}{\sum_f q^2_f}\right](\sum_f a^2_f+\sum_f v^2_f)\nonumber \\
&&+\big(\sum_f v_f\big)^2\left(\frac{3980}{9}-\frac{28960\zeta_3}{9}+\frac{11200\zeta_5}{9}\right) \nonumber \\
&&+\left(\frac{356000}{9}-44192\zeta_3\right)\ln{\frac{M_Z^2}{M_t^2}}+\frac{33776}{3}\ln^2{\frac{M_Z^2}{M_t^2}} \nonumber \\
&&+\frac{8464}{3}\ln^3{\frac{M_Z^2}{M_t^2}}-\frac{13083735979}{56700}+\frac{35934334\zeta_3}{525} \nonumber \\
&&+\frac{12328\zeta_5}{9}-\frac{170272\ln^4{2}}{405}+\frac{512\ln^5{2}}{45} \nonumber \\
&&+\left(11748+\frac{512\ln{2}}{3}+\frac{170272\ln^2{2}}{405}-\frac{512\ln^3{2}}{27}\right)\pi^2\nonumber \\
&&+\left(\frac{4425752}{6075}+\frac{1424\ln{2}}{135}\right)\pi^4, \\
{\hat r}_{4,1}&=& 144\left[\Pi^{NS}_3+\frac{\Pi^S_3\big(\sum_f q_f\big)^2}{\sum_f q^2_f}\right](\sum_f a^2_f+\sum_f v^2_f)\nonumber \\
&&+\big(\sum_f v_f\big)^2\left(\frac{5960}{27}-\frac{1040\zeta_3}{3}-\frac{320\zeta_3^2}{3}+\frac{800\zeta_5}{3}\right), \\
{\hat r}_{4,2}&=& -12\pi^2\gamma^{NS}_2(\sum_f a^2_f+\sum_f v^2_f), \\
{\hat r}_{4,3}&=& -9\pi^2\Pi^{NS}_1(\sum_f a^2_f+\sum_f v^2_f),
\end{eqnarray}
where the expressions for the coefficients $\gamma_i$ and $\Pi_i$ can be found in Refs.\cite{Baikov:2012zm, Baikov:2012zn}.


\begin{thebibliography}{99}

\bibitem{ALEPH:2005ab}
  S.~Schael {\it et al.} [ALEPH and DELPHI and L3 and OPAL and SLD Collaborations and LEP Electroweak Working Group and SLD Electroweak Group and SLD Heavy Flavour Group],
  %Precision electroweak measurements on the $Z$ resonance,
  Phys.\ Rep.\  {\bf 427}, 257 (2006).

\bibitem{Alcaraz:2007ri}
  J.~Alcaraz {\it et al.} [ALEPH and DELPHI and L3 and OPAL Collaborations and LEP Electroweak Working Group],
  %Precision Electroweak Measurements and Constraints on the Standard Model,
  arXiv:0712.0929 [hep-ex].

\bibitem{superZ}
  J.P. Ma and Z.X. Zhang (The super Z-factory group),
  %Z-factory Physics,
  Sci. China: Phys., Mech. Astron. {\bf 53}, 1947 (2010).

\bibitem{CEPCStudyGroup:2018ghi}
  J.~B.~Guimaraes da Costa {\it et al.} [CEPC Study Group],
  %``CEPC Conceptual Design Report: Volume 2 - Physics and Detector,''
  [arXiv:1811.10545 [hep-ex]].

\bibitem{Akhundov:1985fc}
  A.~A.~Akhundov, D.~Y.~Bardin, and T.~Riemann,
  %Electroweak One Loop Corrections to the Decay of the Neutral Vector Boson,
  Nucl.\ Phys.\  B {\bf 276}, 1 (1986).

\bibitem{Novikov:1992rj}
  V.~A.~Novikov, L.~B.~Okun, and M.~I.~Vysotsky,
  %On the Electroweak one loop corrections,
  Nucl.\ Phys.\ B {\bf 397}, 35 (1993).

\bibitem{Novikov:1993ir}
  V.~A.~Novikov, L.~B.~Okun, and M.~I.~Vysotsky,
  %On the electroweak and gluonic corrections to the hadronic width of the Z boson,
  Phys.\ Lett.\ B {\bf 320}, 388 (1994).

\bibitem{Czarnecki:1996ei}
  A.~Czarnecki and J.~H.~Kuhn,
  %Nonfactorizable QCD and electroweak corrections to the hadronic Z boson decay rate,
  Phys.\ Rev.\ Lett.\  {\bf 77}, 3955 (1996).

\bibitem{Freitas:2014hra}
  A.~Freitas,
  %Higher-order electroweak corrections to the partial widths and branching ratios of the Z boson,
  J. High Energy Phys. {\bf 04}, 070 (2014).

\bibitem{Freitas:2013dpa}
  A.~Freitas,
  %Two-loop fermionic electroweak corrections to the Z-boson width and production rate,
  Phys.\ Lett.\ B {\bf 730}, 50 (2014).

\bibitem{Dubovyk:2018rlg}
  I.~Dubovyk, A.~Freitas, J.~Gluza, T.~Riemann, and J.~Usovitsch,
  %Complete electroweak two-loop corrections to Z boson production and decay,
  Phys.\ Lett.\ B {\bf 783}, 86 (2018).

\bibitem{Avdeev:1994db}
  L.~Avdeev, J.~Fleischer, S.~Mikhailov, and O.~Tarasov,
  %$O(\alpha \alpha_s^2)$ correction to the electroweak $\rho$ parameter,
  Phys.\ Lett.\ B {\bf 336}, 560 (1994).

\bibitem{Chetyrkin:1995ix}
  K.~G.~Chetyrkin, J.~H.~Kuhn, and M.~Steinhauser,
  %Corrections of order ${\cal O}(G_F M_t^2 \alpha_s^2)$ to the $\rho$ parameter,
  Phys.\ Lett.\ B {\bf 351}, 331 (1995).

\bibitem{Faisst:2003px}
  M.~Faisst, J.~H.~Kuhn, T.~Seidensticker, and O.~Veretin,
  %Three loop top quark contributions to the $\rho$ parameter,
  Nucl.\ Phys.\ B {\bf 665}, 649 (2003).

\bibitem{vanderBij:2000cg}
  J.~J.~van der Bij, K.~G.~Chetyrkin, M.~Faisst, G.~Jikia, and T.~Seidensticker,
  %Three loop leading top mass contributions to the $\rho$ parameter,
  Phys.\ Lett.\ B {\bf 498}, 156 (2001).

\bibitem{Schroder:2005db}
  Y.~Schroder and M.~Steinhauser,
  %Four-loop singlet contribution to the $\rho$ parameter,
  Phys.\ Lett.\ B {\bf 622}, 124 (2005).

\bibitem{Chetyrkin:2006bj}
  K.~G.~Chetyrkin, M.~Faisst, J.~H.~Kuhn, P.~Maierhofer, and C.~Sturm,
  %Four-Loop QCD Corrections to the $\rho$ Parameter,
  Phys.\ Rev.\ Lett.\  {\bf 97}, 102003 (2006).

\bibitem{Boughezal:2006xk}
  R.~Boughezal and M.~Czakon,
  %Single scale tadpoles and $O(G_F m_t^2 \alpha_s^3)$ corrections to the $\rho$ parameter,
  Nucl.\ Phys.\  B {\bf 755}, 221 (2006).

\bibitem{Kataev:1992dg}
  A.~L.~Kataev,
  %Higher order $O(\alpha^2)$ and $O(\alpha \alpha_s)$ corrections to sigma total $e^+ e^- \rightarrow$ hadron and Z boson decay rate,
  Phys.\ Lett.\ B {\bf 287}, 209 (1992).

\bibitem{Harlander:1997zb}
  R.~Harlander, T.~Seidensticker, and M.~Steinhauser,
  %Complete corrections of Order $\alpha \alpha_s$ to the decay of the Z boson into bottom quarks,
  Phys.\ Lett.\ B {\bf 426}, 125 (1998).

\bibitem{Kniehl:1989qu}
  B.~A.~Kniehl and J.~H.~Kuhn,
  %QCD Corrections to the Z Decay Rate,
  Nucl.\ Phys.\  B {\bf 329}, 547 (1990).

\bibitem{Kniehl:1989bb}
  B.~A.~Kniehl and J.~H.~Kuhn,
  %QCD Corrections to the Axial Part of the Z Decay Rate,
  Phys.\ Lett.\ B {\bf 224}, 229 (1989).

\bibitem{Gorishnii:1990vf}
  S.~G.~Gorishnii, A.~L.~Kataev, and S.~A.~Larin,
  %The $O(\alpha^{3}_{s})$-corrections to $\sigma_{tot}(e^{+}e^{-}\rightarrow hadrons)$ and $\Gamma(\tau^{-} \rightarrow \nu_{\tau} + hadrons)$ in QCD,
  Phys.\ Lett.\ B {\bf 259}, 144 (1991).

\bibitem{Surguladze:1990tg}
  L.~R.~Surguladze and M.~A.~Samuel,
  %Total hadronic cross-section in $e^+ e^-$ annihilation at the four loop level of perturbative QCD,
  Phys.\ Rev.\ Lett.\  {\bf 66}, 560 (1991).

\bibitem{Larin:1993ju}
  S.~A.~Larin, T.~van Ritbergen, and J.~A.~M.~Vermaseren,
  %The $\alpha_s^3$ correction to $\Gamma (Z0 \rightarrow hadrons)$,
  Phys.\ Lett.\ B {\bf 320}, 159 (1994).

\bibitem{Chetyrkin:1993jm}
  K.~G.~Chetyrkin and J.~H.~Kuhn,
  %Complete QCD corrections of order $\alpha_s^2$ to the Z decay rate,
  Phys.\ Lett.\ B {\bf 308}, 127 (1993).

\bibitem{Chetyrkin:1993ug}
  K.~G.~Chetyrkin and O.~V.~Tarasov,
  %The $\alpha_s^3$ corrections to the effective neutral current and to the Z decay rate in the heavy top quark limit,
  Phys.\ Lett.\ B {\bf 327}, 114 (1994).

\bibitem{Baikov:2012xh}
  P.~A.~Baikov, K.~G.~Chetyrkin, J.~H.~Kuhn, and J.~Rittinger,
  %R(s) and Z decay in order $\alpha_s^4$: complete results,
  PoS RADCOR {\bf 2011}, 030 (2011).

\bibitem{Baikov:2008jh}
  P.~A.~Baikov, K.~G.~Chetyrkin, and J.~H.~Kuhn,
  %Order $\alpha^4_s$ QCD Corrections to Z and tau Decays,
  Phys.\ Rev.\ Lett.\  {\bf 101}, 012002 (2008).

\bibitem{Baikov:2012er}
  P.~A.~Baikov, K.~G.~Chetyrkin, J.~H.~Kuhn, and J.~Rittinger,
  %Complete ${\cal O}(\alpha_s^4)$ QCD Corrections to Hadronic $Z$-Decays,
  Phys.\ Rev.\ Lett.\  {\bf 108}, 222003 (2012).

\bibitem{Chetyrkin:1994js}
  K.~G.~Chetyrkin, J.~H.~Kuhn, and A.~Kwiatkowski,
  %QCD corrections to the $e^{+} e^{-}$ cross-section and the $Z$ boson decay rate,
  Phys.\ Rep.\  {\bf 277}, 189 (1996).

\bibitem{Chetyrkin:2000zk}
  K.~G.~Chetyrkin, R.~V.~Harlander, and J.~H.~Kuhn,
  %Quartic mass corrections to $R_{had}$ at $\mathcal O(\alpha^3_s)$,
  Nucl.\ Phys.\ B {\bf 586}, 56 (2000).

\bibitem{Baikov:2004ku}
  P.~A.~Baikov, K.~G.~Chetyrkin, and J.~H.~Kuhn,
  %Vacuum polarization in pQCD: First complete $O(\alpha_s^4)$ result,
  Nucl.\ Phys.\ Proc.\ Suppl.\  {\bf 135}, 243 (2004).

\bibitem{Chetyrkin:1993tt}
  K.~G.~Chetyrkin,
  %Power suppressed heavy quark mass corrections to the $\tau$ lepton and Z boson decay rates,
  Phys.\ Lett.\ B {\bf 307}, 169 (1993).

\bibitem{Larin:1994va}
  S.~A.~Larin, T.~van Ritbergen, and J.~A.~M.~Vermaseren,
  %The Large quark mass expansion of $\Gamma (Z0\rightarrow hadrons)$ and $\Gamma (\tau\rightarrow \tau-neutrino + hadrons)$ in the order $\alpha_s^3$,
  Nucl.\ Phys.\  B {\bf 438}, 278 (1995).

\bibitem{DEnterria:2019lql}
  D.~D'Enterria,
  %High-precision $\alpha_S$ from W and Z hadronic decays,
  PoS ALPHAS {\bf 2019}, 008 (2019).

\bibitem{Brodsky:2012ms}
  S.~J.~Brodsky and X.~G.~Wu,
  %Self-Consistency Requirements of the Renormalization Group for Setting the Renormalization Scale,
  Phys.\ Rev.\ D {\bf 86}, 054018 (2012).

\bibitem{Wu:2013ei}
  X.~G.~Wu, S.~J.~Brodsky, and M.~Mojaza,
  %The Renormalization Scale-Setting Problem in QCD,
  Prog.\ Part.\ Nucl.\ Phys.\  {\bf 72}, 44 (2013).

\bibitem{Wu:2014iba}
  X.~G.~Wu, Y.~Ma, S.~Q.~Wang, H.~B.~Fu, H.~H.~Ma, S.~J.~Brodsky, and M.~Mojaza,
  %Renormalization Group Invariance and Optimal QCD Renormalization Scale-Setting,
  Rep.\ Prog.\ Phys.\  {\bf 78}, 126201 (2015).

\bibitem{GellMann:1954fq}
  M.~Gell-Mann and F.~E.~Low,
  %Quantum electrodynamics at small distances,
  Phys.\ Rev.\  {\bf 95}, 1300 (1954).

\bibitem{Grunberg:1980ja}
  G.~Grunberg,
  %Renormalization Group Improved Perturbative QCD,
  Phys.\ Lett.\ B {\bf 95}, 70 (1980).

\bibitem{Grunberg:1982fw}
  G.~Grunberg,
  %Renormalization Scheme Independent QCD and QED: The Method of Effective Charges,
  Phys.\ Rev.\ D {\bf 29}, 2315 (1984).

\bibitem{Stevenson:1980du}
  P.~M.~Stevenson,
  %Resolution of the Renormalization Scheme Ambiguity in Perturbative QCD,
  Phys.\ Lett.\ B {\bf 100}, 61 (1981).

\bibitem{Stevenson:1981vj}
  P.~M.~Stevenson,
  %Optimized Perturbation Theory,
  Phys.\ Rev.\ D {\bf 23}, 2916 (1981).

\bibitem{Stevenson:1982wn}
  P.~M.~Stevenson,
  %Sense and Nonsense in the Renormalization Scheme Dependence Problem,
  Nucl.\ Phys.\ B {\bf 203}, 472 (1982).

\bibitem{Brodsky:2011ta}
  S.~J.~Brodsky and X.~G.~Wu,
  %Scale Setting Using the Extended Renormalization Group and the Principle of Maximum Conformality: the QCD Coupling Constant at Four Loops,
  Phys.\ Rev.\ D {\bf 85}, 034038 (2012).

\bibitem{Brodsky:2011ig}
  S.~J.~Brodsky and L.~Di Giustino,
  %Setting the Renormalization Scale in QCD: The Principle of Maximum Conformality,
  Phys.\ Rev.\ D {\bf 86}, 085026 (2012).

\bibitem{Brodsky:2012rj}
  S.~J.~Brodsky and X.~G.~Wu,
  %Eliminating the Renormalization Scale Ambiguity for Top-Pair Production Using the Principle of Maximum Conformality,
  Phys.\ Rev.\ Lett.\ {\bf 109}, 042002 (2012).

\bibitem{Mojaza:2012mf}
  M.~Mojaza, S.~J.~Brodsky, and X.~G.~Wu,
  %Systematic All-Orders Method to Eliminate Renormalization-Scale and Scheme Ambiguities in Perturbative QCD,
  Phys.\ Rev.\ Lett.\ {\bf 110}, 192001 (2013).

\bibitem{Brodsky:2013vpa}
  S.~J.~Brodsky, M.~Mojaza, and X.~G.~Wu,
  %Systematic Scale-Setting to All Orders: The Principle of Maximum Conformality and Commensurate Scale Relations,
  Phys.\ Rev.\ D {\bf 89}, 014027 (2014).

\bibitem{Wu:2018cmb}
  X.~G.~Wu, J.~M.~Shen, B.~L.~Du, and S.~J.~Brodsky,
  %Novel demonstration of the renormalization group invariance of the fixed-order predictions using the principle of maximum conformality and the $C$-scheme coupling,
  Phys.\ Rev.\ D {\bf 97}, 094030 (2018).

\bibitem{Wu:2019mky}
  X.~G.~Wu, J.~M.~Shen, B.~L.~Du, X.~D.~Huang, S.~Q.~Wang, and S.~J.~Brodsky,
  %The QCD Renormalization Group Equation and the Elimination of Fixed-Order Scheme-and-Scale Ambiguities Using the Principle of Maximum Conformality,
  Prog.\ Part.\ Nucl.\ Phys.\ {\bf 108}, 103706 (2019).

\bibitem{Boito:2016pwf}
  D.~Boito, M.~Jamin, and R.~Miravitllas,
  %Scheme Variations of the QCD Coupling and Hadronic $\tau$ Decays,
  Phys.\ Rev.\ Lett.\ {\bf 117}, 152001 (2016).

\bibitem{Beneke:1994qe}
  M.~Beneke and V.~M.~Braun,
  %Naive nonAbelianization and resummation of fermion bubble chains,
  Phys.\ Lett.\ B {\bf 348}, 513 (1995).

\bibitem{Neubert:1994vb}
  M.~Neubert,
  %Scale setting in QCD and the momentum flow in Feynman diagrams,
  Phys.\ Rev.\ D {\bf 51}, 5924 (1995).

\bibitem{Beneke:1998ui}
  M.~Beneke,
  %Renormalons,
  Phys.\ Rep.\ {\bf 317}, 1 (1999).

\bibitem{Du:2018dma}
  B.~L.~Du, X.~G.~Wu, J.~M.~Shen, and S.~J.~Brodsky,
  %Extending the Predictive Power of Perturbative QCD,
  Eur.\ Phys.\ J.\ C {\bf 79}, 182 (2019).

\bibitem{Yu:2019mce}
  Q.~Yu, X.~G.~Wu, J.~Zeng, X.~D.~Huang, and H.~M.~Yu,
  %The heavy quarkonium inclusive decays using the principle of maximum conformality,
  Eur.\ Phys.\ J.\ C {\bf 80}, 362 (2020).

\bibitem{Yu:2018hgw}
  Q.~Yu, X.~G.~Wu, S.~Q.~Wang, X.~D.~Huang, J.~M.~Shen, and J.~Zeng,
  %Properties of the decay $H\to\gamma\gamma$ using the approximate $\alpha_s^4$ corrections and the principle of maximum conformality,
  Chin.\ Phys.\ C {\bf 43}, 093102 (2019).

\bibitem{Huang:2020rtx}
  X.~D.~Huang, X.~G.~Wu, J.~Zeng, Q.~Yu, X.~C.~Zheng, and S.~Xu,
  %Determination of the top-quark $\overline{\rm MS}$ running mass via its perturbative relation to the on-shell mass with the help of principle of maximum conformality,
  Phys.\ Rev.\ D {\bf 101}, 114024 (2020).

\bibitem{Yu:2020tri}
  H.~M.~Yu, W.~L.~Sang, X.~D.~Huang, J.~Zeng, X.~G.~Wu, and S.~J.~Brodsky,
  %Scale-Fixed Predictions for $\gamma + \eta_c$ production in electron-positron collisions at NNLO in perturbative QCD,
  J. High Energy Phys.  {\bf 01}, 131 (2021).

\bibitem{Basdevant:1972fe}
  J.~L.~Basdevant,
  %The Pade approximation and its physical applications,
  Fortsch.\ Phys.\ {\bf 20}, 283 (1972).

\bibitem{Samuel:1992qg}
  M.~A.~Samuel, G.~Li, and E.~Steinfelds,
  %Estimating perturbative coefficients in quantum field theory using Pade approximants. 2.,
  Phys.\ Lett.\ B {\bf 323}, 188 (1994).

\bibitem{Samuel:1995jc}
  M.~A.~Samuel, J.~R.~Ellis, and M.~Karliner,
  %Comparison of the Pade approximation method to perturbative QCD calculations,
  Phys.\ Rev.\ Lett.\  {\bf 74}, 4380 (1995).

\bibitem{Shen:2017pdu}
  J.~M.~Shen, X.~G.~Wu, B.~L.~Du, and S.~J.~Brodsky,
  %Novel All-Orders Single-Scale Approach to QCD Renormalization Scale-Setting,
  Phys.\ Rev.\ D {\bf 95}, 094006 (2017).

\bibitem{Zheng:2013uja}
  X.~C.~Zheng, X.~G.~Wu, S.~Q.~Wang, J.~M.~Shen, and Q.~L.~Zhang,
  %Reanalysis of the BFKL Pomeron at the next-to-leading logarithmic accuracy,
  J. High Energy Phys.  {\bf 10}, 117 (2013).

\bibitem{Chawdhry:2019uuv}
  H.~A.~Chawdhry and A.~Mitov,
  %``Ambiguities of the principle of maximum conformality procedure for hadron collider processes,''
  Phys.\ Rev.\ D {\bf 100}, 074013 (2019).

\bibitem{Baikov:2010je}
  P.~A.~Baikov, K.~G.~Chetyrkin, and J.~H.~Kuhn,
  %Adler Function, Bjorken Sum Rule, and the Crewther Relation to Order $\alpha_s^4$ in a General Gauge Theory,
  Phys.\ Rev.\ Lett.\  {\bf 104}, 132004 (2010).

\bibitem{Baikov:2012zn}
  P.~A.~Baikov, K.~G.~Chetyrkin, J.~H.~Kuhn, and J.~Rittinger,
  %Adler Function, Sum Rules and Crewther Relation of Order O($\alpha_s^4$): the Singlet Case,
  Phys.\ Lett.\ B {\bf 714}, 62 (2012).

\bibitem{Bi:2015wea}
  H.~Y.~Bi, X.~G.~Wu, Y.~Ma, H.~H.~Ma, S.~J.~Brodsky, and M.~Mojaza,
  %Degeneracy Relations in QCD and the Equivalence of Two Systematic All-Orders Methods for Setting the Renormalization Scale,
  Phys.\ Lett.\ B {\bf 748}, 13 (2015).

\bibitem{Tanabashi:2018oca}
  M.~Tanabashi {\it et al.} [Particle Data Group],
  %Review of Particle Physics,
  Phys.\ Rev.\ D {\bf 98}, 030001 (2018).

\bibitem{Gross:1973id}
  D.~J.~Gross and F.~Wilczek,
%  ``Ultraviolet Behavior of Nonabelian Gauge Theories,''
  Phys.\ Rev.\ Lett.\  {\bf 30}, 1343 (1973).

\bibitem{Politzer:1973fx}
  H.~D.~Politzer,
%  ``Reliable Perturbative Results for Strong Interactions?,''
  Phys.\ Rev.\ Lett.\  {\bf 30}, 1346 (1973).

\bibitem{Caswell:1974gg}
  W.~E.~Caswell,
%  ``Asymptotic Behavior of Nonabelian Gauge Theories to Two Loop Order,''
  Phys.\ Rev.\ Lett.\  {\bf 33}, 244 (1974).

\bibitem{Tarasov:1980au}
  O.~V.~Tarasov, A.~A.~Vladimirov and A.~Y.~Zharkov,
%  ``The Gell-Mann-Low Function of QCD in the Three Loop Approximation,''
  Phys.\ Lett.\  B {\bf 93}, 429 (1980).

\bibitem{Larin:1993tp}
  S.~A.~Larin and J.~A.~M.~Vermaseren,
%  ``The Three loop QCD Beta function and anomalous dimensions,''
  Phys.\ Lett.\ B {\bf 303}, 334 (1993).

\bibitem{vanRitbergen:1997va}
  T.~van Ritbergen, J.~A.~M.~Vermaseren and S.~A.~Larin,
%  ``The Four loop beta function in quantum chromodynamics,''
  Phys.\ Lett.\ B {\bf 400}, 379 (1997).

\bibitem{Chetyrkin:2004mf}
  K.~G.~Chetyrkin,
%  ``Four-loop renormalization of QCD: Full set of renormalization constants and anomalous dimensions,''
  Nucl.\ Phys.\ B {\bf 710}, 499 (2005).

\bibitem{Czakon:2004bu}
  M.~Czakon,
%  ``The Four-loop QCD beta-function and anomalous dimensions,''
  Nucl.\ Phys.\ B {\bf 710}, 485 (2005).

\bibitem{Baikov:2016tgj}
  P.~A.~Baikov, K.~G.~Chetyrkin and J.~H.~Kuhn,
%  ``Five-Loop Running of the QCD coupling constant,''
  Phys.\ Rev.\ Lett.\  {\bf 118}, 082002 (2017).

\bibitem{Wang:2014aqa}
  S.~Q.~Wang, X.~G.~Wu, and S.~J.~Brodsky,
  %Reanalysis of the Higher Order Perturbative QCD corrections to Hadronic $Z$ Decays using the Principle of Maximum Conformality,
  Phys.\ Rev.\ D {\bf 90}, 037503 (2014).

\bibitem{Abbiendi:2000hu}
  G.~Abbiendi {\it et al.} [OPAL],
  %Precise determination of the Z resonance parameters at LEP: 'Zedometry',
  Eur.\ Phys.\ J.\ C {\bf 19}, 587 (2001).

\bibitem{Abreu:2000mh}
  P.~Abreu {\it et al.} [DELPHI],
  %Cross-sections and leptonic forward backward asymmetries from the Z0 running of LEP,''
  Eur.\ Phys.\ J.\ C {\bf 16}, 371 (2000).

\bibitem{Acciarri:2000ai}
  M.~Acciarri {\it et al.} [L3],
  %Measurements of cross-sections and forward backward asymmetries at the $Z$ resonance and determination of electroweak parameters,
  Eur.\ Phys.\ J.\ C {\bf 16}, 1 (2000).

\bibitem{Barate:1999ce}
  R.~Barate {\it et al.} [ALEPH],
  %Measurement of the Z resonance parameters at LEP,
  Eur.\ Phys.\ J.\ C {\bf 14}, 1 (2000).

\bibitem{Baikov:2012zm}
  P.~A.~Baikov, K.~G.~Chetyrkin, J.~H.~Kuhn, and J.~Rittinger,
  %Vector Correlator in Massless QCD at Order O($\alpha_s^4$) and the QED beta-function at Five Loop,
  J. High Energy Phys. {\bf 07}, 017 (2012).
\end{thebibliography}
\end{document}